
---------------------------------------------------------------------------
\def\KR{Kalb-Ramond}
\def\st{{1\over 2\pi\alpha'}}

\def\p{\Pi_{\mu\nu}}

\def\P{\Pi^{\mu\nu}}

\def\ff{F_{\mu\nu}}

\def\Y{\dot Y}
\def\slf{\Phi_{\mu\nu}}
\def\SLF{\Phi^{\mu\nu}}
\def\J{J^{\mu\nu}}
\def\j{J_{\mu\nu}}
\input phyzzx
\pubnum{UTS-DFT-92-29}
\hfuzz 30pt
\titlepage
\singlespace

\title{Gauge Theory of the String Geodesic Field }
\author{Antonio Aurilia\foot{E-Mail address: AAURILIA@CSUPOMONA.EDU}}
\address{Department of Physics\break
California State Polytechnic University\break Pomona, CA 91768}
\author{Anais Smailagic\foot{E-Mail address: ANAIS@ITSICTP.BITNET}}
\address{International Center for Theoretical Physics, Trieste, Italy}
\andauthor{Euro Spallucci\foot{E-Mail address:
SPALLUCCI@TRIESTE.INFN.IT}}
\address{Dipartimento di Fisica Teorica\break Universit\`a di
Trieste,\break
INFN, Sezione di Trieste\break Trieste, Italy 34014 }
\vskip 2cm
\submit{Phys.Rev.D}
\vfill\eject
\doublespace
\endpage

\abstract

A relativistic string is usually represented by the Nambu-Goto
action in terms of the extremal area of a 2-dimensional timelike
submanifold of Minkowski space. Alternatively, a family of classical
solutions of the string equation of motion can be globally described in terms
of the associated {\it geodesic field.}

In this paper we propose a new gauge theory for the geodesic
field of closed and open strings. Our approach solves the technical and
conceptual
problems affecting previous attempts to describe strings in terms of
local field variables. The connection between the
geodesic field, the string current and the Kalb-Ramond gauge potential is
discussed and clarified.

A non-abelian generalization and the generally covariant form of the model
are also discussed.
\vfill\eject

\REF\Nambu{%
Y.Nambu, Phys.Lett.{\bf 92B}, 327 (1980)
}%
\REF\Hoso{%
Y.Hosotani, Phys.Rev.Lett.{\bf 47}, 399 (1981)
}%
\REF\kasr{%
H.A.Kastrup, Phys.Lett.{\bf 82B}, 237 (1979); Phys.Rep.{\bf 101}, 1 (1983)
\hfill\break
H.A.Kastrup, M.A.Rinke, Phys.Lett.{\bf 105B}, 191 (1981)\hfill\break
M.Rinke, Comm.Math.Phys.{\bf 73}, 265 (1980)\hfill\break
Y.Nambu, Phys.Lett.{\bf 102B}, 149 (1981)
}%
\REF\four{%
A.Aurilia, E.Spallucci, ``~Hamilton-Jacobi formalism and the wave equation
for a relativistic membrane~'', in preparation.
}%
\REF\five{%
A.Aurilia, E.Spallucci, ``~The dual Higgs mechanism and the origin of mass in
the universe~'', submitted to the Gravity Research Foundation for the 1991
essay contest on Gravitation.
}%
\REF\six{%
Ya.B Zel'dovich, JETP Lett.{\bf 6},316 (1967)\hfill\break
A.D. Sakharov, Sov. Phys. Dokl.{\bf12}, 1040 (1968)
}%
\REF\Hosot{%
Y.Hosotani, Phys.Rev.Lett.{\bf 55}, 1719 (1985)
}%
\REF\ker{%
See H.A.Kastrup and M.A.Rinke in ref.3
}%
\REF\nole{%
H.B.Nielsen, P.Olesen, Nucl.Phys.{\bf B57}, 367 (1973)
}%
\REF\kr{%
M.Kalb, P.Ramond, Phys.Rev.{\bf D9}, 2273, (1974)
}%
\REF\pam{%
P.A.M.Dirac, Phys.Rev.{\bf 74}, 817 (1948)
}%
\REF\aashi{%
A.Aurilia, Y.Takahashi, Progr.Theor.Phys.{\bf 66}, n$^o$.2, 693 (1981)
}%
\REF\chth{%
A.Chodos, C.B.Thorn, Nucl. Phys.{\bf B72}, 509, (1974)
}%
\REF\bbhp{%
W.A.Bardeen, I.Bars, A.J.Hanson, R.D.Peccei,\hfill\break
Phys.Rev.{\bf D13}, 2364, (1976)
}%
\REF\bh{%
I.Bars, A.J.Hanson, Phys.Rev.{\bf D13}, 1744, (1976)
}%
\REF\sixten{%
See, for instance, Ruth Gregory, Phys. Rev.{\bf D39}, 2108, (1989) and
various works quoted therein.
}%
\REF\new{%
A.Aurilia, A. Smailagic, E. Spallucci,\hfill\break
Class. Quantum Grav.{\bf 9}, 1883, (1992)
}%
\REF\call{%
G.C.Callan, D.Friedman, E.J.Martinec, M.J.Perry,\hfill\break
Nucl. Phys.{\bf B262}, 593,(1985)
}%
\REF\kik{%
E.S.Fradkin, A.A.Tseytlin, Nucl. Phys.{\bf B261}, 1, (1985)\hfill\break
K.Kikkawa, M.Yamasaki, Progr.Th.Phys.Suppl.{\bf 85}, 228 (1985)
}%
\REF\dual{%
See for example, S.Mandelstam, Phys.Rep.{\bf 13}, 260, (1974)\hfil\break
J.Scherk, Rev.Mod.Phys.{\bf 47}, 123, (1975)
}%
\REF\namb{%
Y.Nambu, Phys.Rev.{\bf D10}, 4262 (1974); Phys. Rep. {\bf23C}, 250 (1956)
}%
\REF\eguchi{%
T.Eguchi, Phys.Lett.{\bf 59B}, 73 (1975)
}%
\REF\rysh{%
S.Ryang, J.Ishida, Progr.Th.Phys.{\bf 66}, 685 (1981)
}%
\REF\wong{%
S.K.Wong, N.Cim.{\bf 65A}, 689, (1970)
}%
\REF\teitel{%
C.Teitelboim, Phys.Lett.{\bf 167B}, 63 (1986)
}%
\REF\freetow{%
D.Z.Freedman, P.K.Townsend, Nucl. Phys.{\bf B177}, 282, (1981)
}%
\REF\smol{%
L.Smolin, Phys.Lett.{\bf 137B}, 379, (1984)
}%
\refsend

\chapter{Introduction, motivations and technical background}

The dynamics of a relativistic point-particle of mass $m$ is encoded into an
action which is essentially the proper length of the particle world-line
$x^\mu=X^\mu(\tau)$, i.e.
$\displaystyle{S=
-m\int d\tau\sqrt{-{\dot X}^\mu{\dot X}_\mu}\equiv\int d\tau L(X,\,{\dot
X})}$,
 where ${\dot X}^\mu(\tau)$ is the tangent 4-vector. Besides its geometrical
meaning, this action also  exhibits
invariance under time reparametrization $\tau\rightarrow\tau'(\tau)$. With
such a
symmetry, this elementary system represents the ``prototype'' of any
theory invariant under general coordinate re-definitions.
The dynamical variables commonly used to
describe point-like particles are either the Hamiltonian pair
$(X^\mu,\,P_\mu)$, where $P_\mu$ is
the {\it linear momentum} conjugate to the tangent vector ${\dot X}^\mu$
according
to $P_\mu=\partial L/\partial{\dot X}^\mu$, or the Lagrangian coordinates
$(X^\mu,\,{\dot X}^\mu)$. The two (~dynamically equivalent~) descriptions
are related to each other by the Legendre transform.

A string is the simplest generalization of a point-like particle: it extends
in one spatial dimension and spans, evolving in time,
a two dimensional world-surface. If ${\cal H}$ is a domain in the space
of the parameters $\displaystyle{\xi^a=(\tau,\sigma)}$ which represent
local
coordinates on the lorentzian string manifold, and $\Omega$ is an
embedding
of  ${\cal H}$ in the Minkowski space $M$, then $\displaystyle{\Omega:
\xi\in
{\cal H}\rightarrow\Omega(\xi) =X^\mu(\xi)\in M}$. In particular, the
tangent
bi-vector in parameter space
$$
{\partial\over\partial\tau}\wedge{\partial\over\partial\sigma}
\eqn\biv
$$
is mapped by $\Omega$ into the tangent bi-vector
$$
{\dot X}^{\mu\nu}={\partial X^\mu\over\partial\tau}\wedge
{\partial X^\nu\over\partial\sigma}
=\delta^{[ab]}\partial_a X^\mu\partial_b X^\nu
\eqn\tg
$$
at each point of the embedded sub-manifold $x^\mu=X^\mu(\xi)$
representing
the string history ${\cal W}$ in Minkowski spacetime.
In the analogy
with the point-particle case, a reparametrization invariant action,
which is proportional to the world-sheet area, can be
assigned to the string according to:
$$
S_{\rm NG}=-\st\int d\tau d\sigma
\sqrt{-{1\over 2}{\dot X}^{\mu\nu}{\dot X}_{\mu\nu}}\equiv
\int d\tau d\sigma L_{\rm NG}(X,\,{\dot X})\ .
\eqn\Hosottion
$$

In the canonical approach based on the analogy with the point-particle case,
one defines {\it two linear momenta}, $P_\mu$ and $P'_\mu$
canonically conjugated to ${\dot X}^\mu=\partial X^\mu/\partial\tau$
and $X^{'\mu}=\partial X^\mu/\partial\sigma$ respectively,
and then one develops the corresponding
hamiltonian formalism. This is the usual starting point to string normal
modes decomposition and subsequent quantization. However, this
approach breaks reparametrization invariance on the world-sheet
from the very beginning, so one of the main features of the model is lost.
In order to preserve this symmetry at all times, one has to
treat $\tau$, $\sigma$ on an equal footing, and this requires a
{\it non-canonical}
formulation of string dynamics, i.e. the introduction of new non-canonical
variables. In this connection, the key remark is that in going from
a point-like object to an extended
system, quantities like $\partial_a X^\mu$ lose their physical meaning of
``~velocities~''; rather, they become {\it projectors} on the string
world-sheet. To maintain the analogy with the point-like particle case
and with the geometric meaning of the velocity as tangent element to the
world-trajectory of the physical object, it is preferable to
define the string velocity as
the tangent bi-vector \tg . In this case, the conjugate dynamical variable
is the {\it area momentum} [\Nambu,\Hoso]
$$
\p\equiv {\partial L_{\rm NG}\over\partial{\dot X}^{\mu\nu}}=
\st{{\dot X}_{\mu\nu}\over\sqrt{-{1\over 2}{\dot X}^{\mu\nu}
{\dot X}_{\mu\nu}}}\ ,
\eqn\amom
$$
which involves {\it both} $P_\mu$ and $P'_\mu$,
\foot{The area momentum
is simply related to the canonical momenta; in fact
$$\eqalign{
\p\partial_c X^\nu&=\st\delta^{[ab]}{\partial_a X_\mu\partial_b
X_\nu\over
\sqrt{-{1\over 2}{\dot X}^{\rho\sigma}{\dot X}_{\rho\sigma}}}
\partial_c X^\nu\cr
&=\st\delta^{[ab]}{\partial_a X_\mu\over
\sqrt{-{1\over 2}{\dot X}^{\rho\sigma}{\dot
X}_{\rho\sigma}}}\gamma_{bc}\cr
&=\st\delta^{[ab]}\gamma_{bc}{\partial_a X_\mu\over
\sqrt{-{1\over 2}{\dot X}^{\rho\sigma}{\dot X}_{\rho\sigma}}}\ ,\cr}
$$
where $\partial_a X_\mu/
\sqrt{-{1\over 2}{\dot X}^{\rho\sigma}{\dot X}_{\rho\sigma}}\equiv
2\pi\alpha'P_{a\,\mu}$, and $P_{0\,\mu}\equiv P_\mu$,
$P_{1\,\mu}\equiv P'_\mu$.}
and, according to eq.(1.4), is proportional to the {\it unit norm
tangent element} to the string world-sheet.
Then, $\Pi_{\mu\nu}$ satisfies the generalized mass-shell condition
$$
-{1\over 2}\p\P={1\over (2\pi\alpha')^2}\
\eqn\masshell
$$
which corresponds to the relativistic particle momentum constraint
$p_\mu p^\mu=-\mu^2$.
Similarly, in terms of the dynamical variables $(X^\mu,\p)$, the string
equations of motion take on the compact form
$$\eqalign{
&\delta^{[ab]}\partial_a\p\partial_b X^\nu=0 \cr
&\left(\p\partial_b X^\nu\right)\big\vert_{\partial{\cal W}}=0\ .\cr}
\eqn\class
$$
The embedding $\Omega$ maps the boundary of ${\cal H}$
into the world-lines of the
the string end-points $x^\mu=X^\mu(\tau,\sigma=\sigma_1(\tau))
=X^\mu_{1)}(\tau)$,
$x^\mu=X^\mu(\tau,\sigma=\sigma_2(\tau))=X^\mu_{2)}(\tau)$.

Thus the first equation in (1.6) represents the conservation  of the area
momentum along the string world-sheet, while the second
provides boundary condition at the string end points.
\medskip
\underbar{\it Motivation and objectives}

To reiterate the main point of our introductory remarks: the choice of
dynamical variables  $(X^\mu,\,\p)$ offers several distinct advantages over
the conventional choice of canonical variables. First, it preserves
covariance on the world-sheet at any
stage in the formulation of string dynamics. This property is particularly
desirable since it suggests a novel approach to the quantum theory of
extended systems [\Hoso]; second, it lends itself to a straightforward
generalization to the case of submanifolds of higher dimensionality and, in
the process, it enlightens the close correspondence between the theory of
extended systems and the Hamilton-Jacobi formulation of the mechanics of
point particles [\kasr,\four].
Against this background, our immediate objective is to show that the new
choice of variables  $(X^\mu,\,\p)$ enables one to cast the first equation of
motion in (1.6)
in the form of a {\it Bianchi Identity}. This property, in turn, opens the
way to the formulation of string
dynamics as the gauge theory of an antisymmetric tensor field. Such a gauge
formulation for strings is the primary purpose of the paper.
In physical terms, the payoff of this new gauge formulation is a
mechanism of {\it mass generation} for antisymmetric tensor fields (~in
this
specific instance, the Kalb-Ramond field~). The resulting equations describe
massive spin-1 particles
and represent the relativistic counterpart of the London equations of
superconductivity. Elsewhere we have speculated that this new mechanism
of
mass generation, when applied to an antisymmetric tensor field of rank-3,
may play an important role in cosmology
in connection with the problem of production of dark matter in the early
universe [\five].
An equally interesting application of the gauge formulation of string
dynamics is briefly discussed in section-3 in connection with the
{\it induced gravity} program pioneered by Zel'dovich and Sakharov [\six] as
a way to get around the long standing problem
of quantizing General Relativity. There we will argue that the Einstein and
Kalb-Ramond terms are generated in the effective action for the
background fields as induced quantum terms describing the low energy
behavior of the underlying quantum string theory.

\underbar{\it Technical background}

Before we embark on a detailed discussion of the gauge formulation
of string dynamics, it may be helpful to address the main technical
difficulty that we need to overcome. Suppose that one is able to invert
the relation between $x^\mu$ and $\xi^a\equiv (\tau,\sigma)$
so that $\partial_a$ in $(1.6)$ is expressed through the chain rule as
$\displaystyle{\partial_a={\partial X^\lambda\over\partial\xi^a}
{\partial\over\partial X^\lambda}}$. Then, the first equation
in (1.6) can be written as
$$
\delta^{[ab]}\partial_a X^\lambda\partial_b X^\nu\partial_\lambda\p=0
\longrightarrow
{\dot X}^{\lambda\nu}\partial_\lambda\p=0\ .
\eqn\motion
$$
However, from the condition (1.5) we deduce that $\P\partial_\alpha\p=0$.
Then, eq.\motion\ can be fully antisymmetrized, yielding
$$
{\dot X}^{\lambda\nu}\partial_{\,[\lambda}\Pi_{\mu\nu]}=0\ .
\eqn\bianchi
$$
Thus, whenever the matrix ${\dot X}^{\lambda\nu}$ is non-degenerate,
eq.\bianchi\ implies that
$\p$ satisfies a Bianchi-type Identity and therefore can be written, at least
locally, in terms of a gauge potential $B_\mu(x)$.
{\it The above remarks are quite
general and apply both to closed and open strings.}

Unfortunately, this result hinges on two assumptions which are at least
questionable. In fact:

\noindent
i) in order to invert the relation $x^\mu=X^\mu(\tau,\,\sigma)$ one has to
consider a two-parameter family of classical solutions
$x^\mu=X^\mu(\tau,\sigma,\phi_1,\phi_2)$ and assign to $\phi_1,\,\phi_2$
the role of additional coordinates in parameter space [\Hosot].
Then, the embedding functions
establish a mapping between spacetime and parameter space which,
however,
may not be one-to-one and is thus beset with integrability problems. In any
case,
this kind of approach which describes strings in
terms of a pair of scalar fields, i.e. $\phi_1$,$\phi_2$, is hard to
interpret as a
genuine gauge theory.

\noindent
ii) If the matrix ${\dot X}^{\mu\nu}$ represents the tangent element to the
string world-sheet, then it is degenerate. In fact
$$\eqalign{
{\rm det}{\dot X}^{\mu\nu}&=\epsilon_{\mu\nu\rho\sigma}
\epsilon_{\mu'\nu'\rho'\sigma'}
{\dot X}^{\mu\mu'}{\dot X}^{\nu\nu'}{\dot X}^{\rho\rho'}
{\dot X}^{\sigma\sigma'}\cr
&=\delta^{[ab]}\delta^{[cd]}\delta^{[ef]}\delta^{[gh]}
\epsilon_{\mu\nu\rho\sigma}\partial_a X^\mu\partial_c X^\nu\partial_e
X^\rho
\partial_g X^\sigma
\epsilon_{\mu'\nu'\rho'\sigma'}\partial_b X^{\mu'}\partial_d X^{\nu'}
\partial_f X^{\rho'}\partial_h X^{\sigma'}\cr
&\equiv 0\ ,\cr}
\eqn\degen
$$
since any four index totally antisymmetric tensor
in two dimensions is identically zero.   Therefore, eq.\bianchi\
{\it does not imply the existence of
a gauge potential} for the string, even locally.

With the above observations in mind, in the next sections we shall
discuss a lagrangian field theory
which provides a {\it consistent} description of string dynamics
in terms of local
gauge fields. Presently, we shall briefly introduce the formal apparatus
required to realize this programme.

The mathematical object needed
to give $\Pi_{\mu\nu}(\xi)$ the status of local variable, i.e. defined
at any spacetime point rather than on the string world-sheet alone,
is the {\it slope field} or sheet field [\kasr,\Hoso] $\SLF(x)$.
The slope field is a {\it totally anti-symmetric tensor which assigns a
tangent plane at any spacetime point.} More precisely,
due to its geometrical meaning, the slope-field is
characterized by the following properties:

\noindent
a) when evaluated on the string world-sheet, the slope field coincides with
the tangent element, i.e. $\displaystyle{\SLF\left(x=X\right)=
{\dot X}^{\mu\nu}}$;

\noindent
b) it is orthogonal to its dual, i.e.
$\displaystyle{\widetilde\SLF\slf=0}$\foot{Sometimes in the literature
this
property is referred to as the Pl\"ucker condition [\kasr].};

\noindent
c) if it satisfies the Bianchi Identities, i.e.
$\displaystyle{\partial_{\,[\lambda}\Phi_{\mu\nu]}=0}$, then it is called
{\it geodesic field} [\kasr].

Properties b) and c) imply that the 2-form $\Phi\equiv{1\over
2}\slf(x)dx^\mu
\wedge dx^\nu$ has rank 2 in four dimensions, rather than four, i.e. there
exist a coordinate system where $\slf$ can be written in terms of a pair
of Clebsch [\Nambu]
potentials $S^{1)}(x)$ and $S^{2)}(x)$ as $\Phi(x)=dS^{1)}\wedge
dS^{2)}$. This property is essential to solve the problem i) [\ker].

The role of the slope field in connection with string dynamics has already
been discussed by some authors, but mainly from a kinematical point of
view as a useful device to describe a family of minimal surfaces solving the
classical string equation of motion [\Nambu-\kasr].

Property c) led Nielsen and Olesen [\nole] to identify the string field
strength
as the dual of a closed world-sheet
$$
F_{\mu\nu}={1\over 2}\epsilon_{\mu\nu\rho\sigma}\int_{D:\partial
D=\emptyset}
d^2\sigma \,{\dot X}^{\rho\sigma}\,\delta^{4)}\left[x-X(\sigma)\right]
\equiv {1\over 2}\epsilon_{\mu\nu\rho\sigma}J^{\rho\sigma}\ .
\eqn\nielsen
$$
Here, the absence of a boundary guarantees that $F_{\mu\nu}$ defined by
\nielsen\
satisfies the Bianchi Identity so that, at least locally, a gauge potential can
be defined:
$$
\epsilon^{\alpha\beta\mu\nu}\partial_\beta F_{\mu\nu}=
2\partial_\mu J^{\mu\alpha}
=0\,\Rightarrow\, F_{\mu\nu}=\partial_{\,[\mu}A_{\nu]}\ .
\eqn\potenziale
$$
Therefore, in the Nielsen Olesen formulation, $F_{\mu\nu}$ acquires the
meaning of string geodesic field. However,
the ansatz \nielsen\ has several drawbacks: first, it appears that only
closed strings can be given a gauge type description; second, it is not
possible to prove that \nielsen\ is a solution of the field equations of
$A_\mu$; third, the degeneracy of the matrix ${\dot X}^{\mu\nu}$ forbids
a frame-independent
derivation of the string equation of motion from the $A_\mu$ field equation.

Our main purpose is to derive a {\it general gauge description both for
closed and open strings, solving the above technical and conceptual
problems}.
The way-out is to give up the ansatz \nielsen\ and {\it to consider the slope
and the geodesic fields as basically distinct objects, which are related to
each other by a set of classical field equations derived from a
suitable Lagrangian density}. The slope field accounts for the geometric
properties
a) and b); instead, the geodesic field represents the gauge partner of the
slope field,
and satisfies the Bianchi Identity by definition. Then, the solutions of the
field
equations provides the link between them, and relates geometric features
to gauge properties of the string. However, in order to implement this
program, one has
 to revise, from the very beginning, the relation between the
slope field and the string current; then, one has to discuss the connection
between the slope field, the string current and the
Kalb-Ramond potential which mediates the gauge interaction between string
elements.

Basically, the slope field is nothing but the generalization of the
notion of {\it velocity field} in a continuous medium. Suppose that a
spacetime
region is completely filled with a fluid of point-like particles,
each moving along
non-intersecting world-lines. Rather than describing the fluid dynamics in
terms of the motion of each microscopic constituent, one defines a regular
vector field which matches
at any point the 4-velocity of the corresponding particle.
This idea can be applied to the string theory as well, in which case $\SLF$
can be related to the string current
$$
J^{\mu\nu}(x)=\int_{\cal H} d^2\xi\,\delta^{4)}\left(x-Y(\xi)\right)
\Y^{\mu\nu}\ .
\eqn\curr
$$
Indeed
$$\eqalign{
J^{\mu\nu}(x=X)&=
\int d^2\xi'\delta^{4)}\left(X(\xi)-Y(\xi')\right)\dot Y^{\mu\nu}\cr
&={\rm const.}{1\over a^2}\int d^2\xi'\delta^{2)}\left(\xi-\xi'\right)
{\dot X^{\mu\nu}\over\sqrt{-{1\over2}{\dot X}_{\rho\sigma}
{\dot X}^{\rho\sigma}}}\cr
&={\rm const.}'\Pi^{\mu\nu}(\xi)\cr}
\eqn\currpi
$$
where we have explicitly regularized the distribution $\J$ by assigning a
physical width $a$ to the string. Notice that the string current
satisfies the condition
$$\eqalign{
&\widetilde J^{\mu\nu}(x)J_{\mu\nu}(x')=\cr
&\int d^2\xi\int d^2\xi'
\delta^{4)}\left(x-X(\xi)\right)\delta^{4)}\left(x-Y(\xi')\right)
\delta^{[ab]}\delta^{[cd]}\epsilon_{\mu\nu\rho\sigma}\partial_a X^\mu
\partial_b X^\nu\partial_c Y^\rho\partial_d Y^\sigma\equiv 0\cr}
\eqn\pluck
$$
again, because there is no totally antisymmetric four index tensor in two
dimensions.
The only relevant difference between the string current and the slope field
is that $J^{\mu\nu}(x)$ is a distribution different from zero only along
the string world-sheet, while $\SLF(x)$ is a field defined over the whole
spacetime manifold. The above results are concisely expressed by the
relationship
$$
J^{\mu\nu}(x)=\SLF(x)\int d^2\xi\,\delta^{4)}\left(x-X(\xi)\right)\ ,
\eqn\jsl
$$
which is {\it our own definition of the slope field in terms of the
string current}.
By comparing eq.\curr\ with eq.\jsl\ one would be tempted to say that
$\SLF(x)$ is
nothing but ${\dot Y}^{\mu\nu}$ with $Y(\xi)$ replaced by $x^\mu$ [\nole].
This identification, however, must be made with caution. As a matter of
fact, the geodesic field resulting
from a given family of minimal surfaces which are solutions of the
classical
string
equation of motion, should be constructed according to the following
procedure: given a classical
solution $X^\mu(\xi)$, compute the corresponding area momentum, then
use both the embedding equations $x^\mu=X^\mu(\xi)$ and the string
equation
of motion to write the area momentum as a function of $x$. Then, the
resulting
field
is a smooth function of the coordinates and represents the ``~canonical~''
extension of $\Pi_{\mu\nu}$ in the sense that the 2-form
$\Pi(x)\equiv {1\over 2}\Pi_{\mu\nu}dx^\mu\wedge dx^\nu$ has rank
two [\kasr].

The rest of the paper is organized as follows:

in Sect.2 we study a non-linear
Lagrangian for the geodesic field of both open and closed strings interacting
with Kalb-Ramond and electromagnetic potentials. The on-shell equivalence
of this model with ordinary string theory is shown.

In Sect.3 we discuss the coupling of the string geodesic field to gravity, and
its non-abelian generalization.

\def\sc{{\cal S}^{\rm cl.}}
\def\so{{\cal S}^{\rm op.}}
\def\w{W_{\mu\nu}}
\def\W{W^{\mu\nu}}
\def\b{B_\mu}

\def\j{J_{\mu\nu}}
\def\J{J^{\mu\nu}}

\chapter{The String Gauge Field Strength}

In this section we shall introduce a non-linear Lagrangian for the string
geodesic field which is partly suggested by the physical interpretation
of strings
as {\it extended solitons} of an underlying local field theory, and partly
by earlier investigation of string non-linear electrodynamics
by Nambu [\Nambu,\kasr], Nielsen and Olesen [\nole].

The two cases of open and closed strings have to be discussed
separately.

\underbar{\it Closed strings}

Let us consider the following action
$$\eqalign{
& \sc=-{\bar g}^2\int d^4x\,\sqrt{-{1\over 2}\w\W}+
{1\over 2}\int d^4x\,\W\partial_{\,[\mu}B_{\nu]} \cr
& \ff(x)\equiv\partial_{\,[\mu}B_{\nu]}(x)\ ,\cr}
\eqn\mine
$$
where $\w(x)$ is a totally antisymmetric tensor and $\bar g$ is a
dimensional
constant. Physical dimensions are assigned as follows:
$[\w]=[\ff]=[{\bar g}^2]=
{\rm M}^2$. An action of this type was proposed by Nambu as
an {\it effective abelian theory} interpolating between QCD and classical
string dynamics [\kasr].

At this stage, the $B$-field appearing in eq.(2.1) is simply a Lagrange
multiplier enforcing
a ``~transversality~'' condition for the $\w$ field; we shall see that
on-shell the $B_\mu$ -field becomes the {\it string gauge potential}.
In fact, by
varying the action $(2.1)$ with respect to $\b$ and $\w$, we get the
following
set of field equations
$$
\partial_\mu\W=0\ ,
\eqn\diw
$$
$$
{\bar g}^2{\w\over\sqrt{-{1\over 2}W_{\alpha\beta}W^{\alpha\beta}}}
+F_{\mu\nu}=0\ .
\eqn\feq
$$
The {\it closed} string appears as a special solution of \diw\ , namely
$$\eqalign{
&\hat\W(x)=c\int_{\cal H}d^2\xi\,\delta^{4)}\left(x-
X(\xi)\right){\dot X}^{\mu\nu}=
c\,J^{\mu\nu}(x)\ ,\cr
&c=\hbox{dimensionless const.}\ .\cr}\eqn\sol
$$
The general solution of \diw, i.e.,
$\displaystyle{\W=cJ^{\mu\nu}+\epsilon^{\mu\nu\rho\sigma}\partial_\rho}
V_\sigma$, includes a ``~radiation part~'' described by
a vector field $V_\lambda(x)$, which we set equal to zero everywhere
in the following discussion since our present purpose is to identify
string-like solutions of the action (2.1).

The right hand side of \sol\ is, except for a multiplicative constant,
the current distribution associated with the two-dimensional manifold
${\cal W}$ representing the string history. If the string is spatially
closed, then $\partial\cal W=\emptyset$ and $\J$ has {\it vanishing
divergence}.
Eq.\feq\ is an algebraic relation linking the slope field $\W(x)$
to the string field strength $F_{\mu\nu}$:
$$
F_{\mu\nu}(x)=-{\bar g}^2{\hat\w(x)\over\sqrt{-{1\over 2}\hat
W_{\alpha\beta}
\hat W^{\alpha\beta}}}=-{\bar g}^2
{\j(x)\over\sqrt{-{1\over 2}J_{\alpha\beta}J^{\alpha\beta}}}
=-{\bar g}^2{\Phi_{\mu\nu}(x)\over\sqrt{-{1\over 2}\Phi_{\alpha\beta}
\Phi^{\alpha\beta}}}
\eqn\fsol
$$
from which it follows that the Hamilton-Jacobi (H-J) equation
$$
-{1\over 2}\ff(x) F^{\mu\nu}(x)={\bar g}^4
\eqn\ffield
$$
holds {\it everywhere}.
Thus, $F={1\over 2}\ff\,dx^\mu\wedge dx^\nu$
(which is {\it closed} by definition)
is a 2-form which satisfies (on-shell) the generalized H-J equation \ffield .

The net result of these manipulations is
that while $\w(x)$ is a {\it singular} field having support only
along the string history, $F_{\mu\nu}$ is defined
{\it over the whole spacetime manifold. } Furthermore, when evaluated
on the string world-tube, $F_{\mu\nu}$ is proportional to the
area conjugate momentum. In fact,
$$
F_{\mu\nu}\left(x=X(\xi)\right)=
-{\bar g}^2{{\dot X}_{\mu\nu}\over\sqrt{-{1\over 2}{\dot X}_{\alpha\beta}
{\dot X}^{\alpha\beta}}}\equiv -{1\over c}\Pi_{\mu\nu}\ .
\eqn\fpi
$$
Conversely, eq.\fsol\ defines the {\it area
field} $\Pi_{\mu\nu}(x)$ which is the canonical(=rank-2) extension of the
volume momentum:
$$
\Pi_{\mu\nu}(x)\equiv -c\ff(x)=
\st{\Phi_{\mu\nu}(x)\over\sqrt{-{1\over 2}\Phi_{\alpha\beta}
\Phi^{\alpha\beta}}}\ .
\eqn\volf
$$
Note that eq.\fpi\ implies that we identify the term
${\bar g}^2c$ with
the string tension. That this is indeed the case can be verified directly
by inserting the solution $(2.4)$ into the action $(2.1)$. This operation
yields the classical effective action for $X(\xi)$,
$$\eqalign{
S^{\rm eff.}&=-c{\bar g}^2\int d^4x\sqrt{-{1\over 2}\Phi_{\alpha\beta}
\Phi^{\alpha\beta}
\left(\,\int_{\cal H}d^3\xi\,\delta^{4)}\left(x-X(\xi)\right)\right)^2}\cr
&=-c{\bar g}^2\int d^4x\int_{\cal H}d^2\xi\, \delta^{4)}\left(x-X(\xi)\right)
\sqrt{-{1\over 2}{\dot X}_{\alpha\beta}{\dot X}^{\alpha\beta}}\cr
&=-\st\int_{\cal H}d^2\xi
\sqrt{-{1\over 2}{\dot X}_{\alpha\beta}{\dot X}^{\alpha\beta}}\ .\cr}
\eqn\seff
$$
which represents the action for a free string with an
{\it effective string tension} $1/2\pi\alpha'\equiv c{\bar g}^2$.\foot{
Note that second term in $(2.1)$ does not contribute to \seff\ since
$\partial_\mu \J=0$}

Finally, as a consistency check, we wish to show that the gauge field
representation of the string in terms of $F_{\mu\nu}$ leads to the classical
equations of motion \class .
To this end, we recall that $F_{\mu\nu}$ satisfies the Bianchi identities
everywhere, so that in view of eq. (2.8)
$$
\partial_{\,[\lambda}\Pi_{\mu\nu]}(x)=0
\eqn\bbianchi
$$
at each spacetime point.
Then we can project eq.\bbianchi\ along the string history, that is, we
evaluate $\Pi_{\mu\nu}(x)$ at $x=X(\xi)$ and take the interior product
with ${\dot X}^{\lambda\mu}$:
$$
\eqalign{
&{\dot X}^{\lambda\mu}\partial_{\,[\lambda}\Pi_{\mu\nu]}(\xi)=\cr
&\delta^{[ab]}\partial_a X^\lambda\partial_b X^\mu
\partial_{\,[\lambda}\Pi_{\mu\nu]}(\xi)=\cr
&\delta^{[ab]}\partial_b X^\mu
\partial_a\Pi_{\mu\nu}(\xi)=0\ .\cr}
\eqn\check
$$
The last line in \check\ is just the classical equation of motion
\class\  of the string.
\medskip
\underbar{\it The gauge interaction}

Closed inter-string interaction is known to be mediated by the Kalb-Ramond
gauge potential $A_{\mu\nu}(x)$ [\kr]. Thus, it seems natural to ask what is
the
relationship, if any, between the geodesic field associated to the string
and its gauge partner $A_{\mu\nu}(x)$. In other words, once it is accepted
that the action \mine\ describes a theory of closed strings, the next step
is to study how to introduce the interaction with the Kalb-Ramond field.

Let us consider the following model:
$$
\eqalign{
\sc=\int d^4x\Biggl[&-{\bar g}\sqrt{-{1\over 2}\w\W}+
{1\over 2}\W\partial_{\,[\mu}B_{\nu]} \cr
&+{\kappa\over 2}\W A_{\mu\nu}-{1\over 2\cdot
3!}H_{\mu\nu\rho}H^{\mu\nu\rho}
\Biggr]\cr}
\eqn\actg
$$
where the coupling constant $k$ has the dimension of mass, and
$H_{\mu\nu\rho}=
\partial_{\,[\mu}A_{\nu\rho]}$ is the Kalb-Ramond field strength.
The action \actg\ is invariant under the set of transformations
$$\eqalign{
&\delta A_{\mu\nu}=-{1\over \kappa}\partial_{\,[\mu}\Lambda_{\nu]}\ ,\cr
&\delta B_\mu =\Lambda_\mu+\partial_\mu\phi\ ,\cr
&\delta \W=0\ ,\cr}
\eqn\gt
$$
which shows that under the generalized gauge transformation of the
Kalb-Ramond potential, the $B$-field transforms as the corresponding
{\it Goldstone Boson} . From this viewpoint, $B_\mu$ can be seen as the
gauge part of $A_{\mu\nu}$ or as a Stueckelberg compensating field.

Now the field equations become
$$
\partial_\mu\W=0\ , \eqn\dew
$$
$$
{\bar g}^2{\w\over\sqrt{-{1\over 2}W_{\alpha\beta}W^{\alpha\beta}}}+
\partial_{\,[\mu}B_{\nu]} +\kappa A_{\mu\nu}=0\ ,
\eqn\amn
$$
$$
\partial_\mu H^{\mu\nu\rho}+\kappa W^{\nu\rho}=0\ .\eqn\deh
$$
Again, the closed string appears as a special solution of the type \sol\ of
eq.\dew , while the field strength of the string, i.e. the geodesic field
of the string, can be absorbed into a redefinition of the Kalb-Ramond field
$\tilde A_{\mu\nu}$ which is gauge invariant under \gt ,
and therefore can be interpreted as the physical string field strength
$$
\kappa\tilde A_{\mu\nu}\equiv
\kappa A_{\mu\nu}+\partial_{\,[\mu}B_{\nu]}=-{\bar g}^2
{\w\over\sqrt{-{1\over 2}W_{\alpha\beta}W^{\alpha\beta}}}\ , \eqn\asol
$$
$$
-{1\over 2}\tilde A_{\mu\nu}\tilde A^{\mu\nu}={\bar g^4\over \kappa^2}\ ,
\eqn\aconstr
$$
$$
\tilde A_{\mu\nu}(x=X(\xi))=-{{\bar g}^2\over \kappa}
{{\dot X}_{\mu\nu}\over
\sqrt{-{1\over 2}{\dot X}_{\alpha\beta}{\dot X}^{\alpha\beta}}}
\equiv -{1\over ck}\Pi_{\mu\nu}(\xi)\ .\eqn\api
$$
If we insert the solutions of eqs.(2.14-2.16) in the action
\actg , and take into account that the
the field strength $H_{\mu\nu\rho}$ of $A_{\mu\nu}$ and
$\tilde H_{\mu\nu\rho}$ of $\tilde A_{\mu\nu}$ are the same, then
we find the effective action
$$
S^{\rm eff.}=-\st\int_{\cal H}d^2\xi
\sqrt{-{1\over 2}{\dot X}_{\alpha\beta}{\dot X}^{\alpha\beta}}
+{\bar\kappa\over 2}\int d^4x
J^{\mu\nu}\tilde A_{\mu\nu}-{1\over 2\cdot 3!}\int d^4x\tilde
H_{\mu\nu\rho}
\tilde H^{\mu\nu\rho} \eqn\kract
$$
which is just the usual Kalb-Ramond classical action coupled to a closed
string, where, $\displaystyle{1/2\pi\alpha'\equiv{c{\bar g}^2}}$
is the effective string tension, and
$\displaystyle{\bar \kappa\equiv c\kappa}$ is the effective coupling
constant.

By inverting eq.\api\  one obtains for the string geodesic field
$\displaystyle{\Pi_{\mu\nu}(x)=-{\bar\kappa}\tilde A_{\mu\nu}(x)}$. Then,
the curl of $\Pi_{\mu\nu}(x)$ turns out to be proportional to the Kalb-
Ramond
field strength
$$
\partial_{\,[\mu}\Pi_{\nu\rho]}(x)=-{\bar\kappa}\tilde H_{\mu\nu\rho}(x)\
\eqn\curl
$$
Finally, by projecting eq.\curl\ on the string-world sheet, we obtain
$$
{\dot X}^{\mu\nu}\partial_{\,[\mu}\Pi_{\nu\rho]}(x)=
-{\bar\kappa}\tilde H_{\mu\nu\rho}(x)
{\dot X}^{\mu\nu}\Rightarrow \delta^{[ab]}\partial_a\p\partial_b X^\nu=
-{\bar\kappa\over 2}\tilde H_{\rho\nu\mu}(x){\dot X}^{\rho\nu}\ ,
\eqn\lorentz
$$
which is the ``~Lorentz force~'' equation for the string. Furthermore, by
substituting
the solution \sol\ into eq.\deh , we obtain the Kalb-Ramond
field equation coupled to the string current
$$
\partial_\mu H^{\mu\nu\rho}={\bar\kappa}J^{\nu\rho}\ .
\eqn\kreq
$$

\underbar{\it Open string}

The open case requires some further discussion. In fact, the current of an
open string has a non vanishing divergence, and thus it does not
satisfy eq.$(2.2)$. More precisely
$$
\partial_\mu J^{\mu\nu}=
\int d\tau \sum_{i=1,2}(-1)^i\delta^{4)}\left(x-
X(\tau,\sigma=\sigma_i(\tau))
\right)
\left({dX^\nu\over d\tau}\right)_{\sigma=\sigma_i}
\equiv J^\nu(x)
\eqn\divj
$$
where $\displaystyle{x^\mu=X^\mu(\tau,\sigma=\sigma_i(\tau))}$ represent
the
world-lines of the two string end points $\sigma=\sigma_1,\,\,
\sigma=\sigma_2$. Read from right to left, eq.\divj\ represents the
old ``~trick~'' used by Dirac to describe the electrodynamics
of a pair of opposite point-charges in terms of string variables [\pam].
The boundary current $J^\nu$ has vanishing divergence as consequence
of the identity $\partial_\mu \partial_\nu J^{\mu\nu}\equiv 0$. The above
remarks suggest the rationale to modify
our gauge field formalism in order to be able to describe the open
string as well. The action \actg\ has to be supplemented with terms
describing the motion of the string end points and with an abelian vector
gauge potential
to compensate for the ``~leakage~'' of symmetry through the boundary:
$$\eqalign{
\so=&\int d^4x\Biggl[-{\bar g}^2\sqrt{-{1\over 3!}\w(x)\W(x)}+
{1\over 2}\W(x)\partial_{\,[\mu}B_{\nu]}(x)
+fB_\mu(x)J^\mu(x)\cr
&+{\kappa\over 2}\W(x)A_{\mu\nu}-{1\over 2\cdot
3!}H_{\mu\nu\rho}H^{\mu\nu\rho}
\cr
&+eA_\mu J^\mu-{1\over 4}\left(F_{\mu\nu}-{f\kappa\over
e}A_{\mu\nu}\right)
\left(F^{\mu\nu}-{f\kappa\over e}A^{\mu\nu}\right)\Biggr]\cr
&-\mu_0\int d\tau\sum_{i=1,2}\sqrt{-{\dot X}_{i)}{}^\mu{\dot X}_{i)\,\mu}}
\ ,\cr}
\eqn\openact
$$
where $\kappa$ is the Kalb-Ramond coupling constant, and $\mu_0$ is the
mass of the particles located at the string boundaries.
Note that the action depends now explicitly on $B$ because
of the coupling to the boundary current, whereas the action $\actg$
depends on $B$  only through its field strength. With hindsight one
realizes that the action (2.25) is designed to ensure that a special
solution of the type
\sol\ still exists.
The model is now invariant
under the {\it extended gauge transformations}
$$\eqalign{
&\delta A_{\mu\nu}=\partial_{\,[\mu}\Lambda_{\nu]}\ ,\cr
&\delta B_\mu =\partial_\mu\theta-\kappa\Lambda_\mu\ ,\cr
&\delta A_\mu=\partial_\mu\phi+{\kappa f\over e}\Lambda_\mu\ ,\cr
&\delta W^{\mu\nu}=0\ ,\cr
&\delta X^\mu{}_{i)}(\tau)=0\ .\cr}
\eqn\extgt
$$
{}From the above gauge transformations it follows that
$\displaystyle{\delta H^{\mu\nu\rho}=0}$, but
$$
\delta F_{\mu\nu}={\kappa f\over e}\partial_{\,[\mu}\Lambda_{\nu]}\
\eqn\fgvar
$$
so that F does not represent   a physical quantity. Gauge invariant field
strengths can be assembled as follows:
$$\eqalign{
&\delta\left(F_{\mu\nu}-{f\kappa\over e}A_{\mu\nu}\right)=0\ ,\cr
&\delta\left(eF_{\mu\nu}+f\partial_{\,[\mu}B_{\nu]}\right)=0\ ,\cr
&\delta\left(\partial_{\,[\mu}B_{\nu]}+\kappa A_{\mu\nu}\right)=0\ .\cr}
\eqn\strengths
$$
The set of eqs. \strengths\ displays the mixing of various fields to form
physical(~=gauge invariant~) quantities; specifically, the first equation in
\strengths\ justifies, {\it a posteriori}, the non-standard
choice for the $A_\mu$ kinetic term in \openact.
Varying the action \openact\  with respect to $A_{\mu\nu}$,
$A_\mu$, $\b$, $\w$ and $X^\mu(\tau)$
we get the following set of field equations
$$
\delta_{A_{\mu\nu}}\so=0:\qquad
\partial_\mu H^{\mu\nu\rho}-
{f\kappa\over e}\left(F^{\nu\rho}-{f\kappa\over e}A^{\nu\rho}\right)
=-\kappa W^{\nu\rho}\ ,\eqn\dhw
$$
$$
\delta_{B_{\mu}}\so=0:\qquad \partial_\mu\W(x)=fJ^\nu(x)\ , \eqn\dwj
$$
$$
\delta_{W_{\mu\nu}}\so=0:\qquad
{\bar g}^2{\w(x)\over\sqrt{-{1\over 2}W_{\alpha\beta}W^{\alpha\beta}}}
+\partial_{\,[\mu}B_{\nu]}+\kappa A_{\mu\nu}=0\ ,\eqn\abw
$$
$$
\delta_{X_{i)}{}^{\mu}(\tau)}\so=0:\qquad
{dP_{i)\,\mu}\over d\tau}=
(-1)^i\left(f\partial_{\,[\mu}B_{\nu]}+eF_{\mu\nu}\right)
{\dot X}_{i)}{}^\nu\ ,  \eqn\pdot
$$
$$
\delta_{A_{\mu}}\so=0:\qquad
\partial_\mu \left(F^{\mu\nu}-{f\kappa\over e}A^{\mu\nu}\right)=eJ^\nu
\ ; \eqn\max
$$
where
$$
P_{i)\,\mu}=-\mu_0{\dot X_{i)\,\mu}\over
\sqrt{-{\dot X}_{i)}{}^\nu{\dot X}_{i)\,\nu}}}\ .
\eqn\pimu
$$
At this point several comments seem appropriate: a) the coupled field
equations above involve only gauge invariant combinations
of the various field variables.
Eq.\abw\ is the same as \amn , and again relates $A_{\mu\nu}$ and
$B_{\mu\nu}$ to $\w$; as promised,
eq.\dwj\ admits a special solution $\hat\W$ which is proportional to
the current of an {\it open string} having the two world-lines
$x=X_{1)}(\tau)$
and $x=X_{2)}(\tau)$ as its only boundary:
$$\eqalign{
&\hat\W(x)=fJ^{\mu\nu}(x)\ ,\cr
&J^{\mu\nu}(x)=
\int_{\cal H}d\tau d\sigma\,\delta^{4)}\left(x-X(\tau,\sigma)\right)
{\dot X}^{\mu\nu}\ ,\quad\partial_\mu J^{\mu\nu}(x)=J^\nu(x)\ .\cr}
\eqn\solwj
$$

b) Equation\pdot\ describes the motion of the boundary under the combined
action of the string geodesic field  $\partial_{\,[\mu}B_{\nu]}$ and the
Lorentz force acting on the charged end points.

c) Equation \abw\ relates $B_\mu$ and $A_{\mu\nu}$ to the slope field :
$$
\partial_{\,[\mu}B_{\nu]}+\kappa A_{\mu\nu}=
-{\bar g}^2{\Phi_{\mu\nu}(x)\over\sqrt{-{1\over 3!}\Phi_{\alpha\beta}
\Phi^{\alpha\beta}}}
=-{1\over f}\Pi_{\mu\nu}(x)\ ,
\eqn\abpi
$$
where the effective string tension now is $1/2\pi\alpha'\equiv {\bar g}^2f$.

d) Using the string solution \solwj\ ,
the above system of field equations can be written in the form:
$$
\partial_\mu H^{\mu\nu\rho}-
{f\kappa\over e}\left(F^{\nu\rho}-{f\kappa\over e}A^{\nu\rho}\right)
=-f\kappa J^{\nu\rho}\ , \eqn\dhj
$$
$$
{dP_{i)\,\mu}\over d\tau}=
(-1)^i\left(\Pi_{\mu\nu}-f\kappa A_{\mu\nu}+eF_{\mu\nu}\right){\dot
X}_{i)}
{}^\nu
\qquad i=1,2 \eqn\loren
$$
$$
\partial_\nu\left(F^{\nu\rho}-{f\kappa\over e}A^{\nu\rho}\right)
=eJ^\rho  \eqn\dfj\ ,
$$
$$
\partial_{\,[\mu}\Pi_{\nu\rho]}=-f\kappa H_{\mu\nu\rho}\ . \eqn\curlpi
$$

The interpretation of the above equations is as follows. First,
equation \curlpi\ which is the covariant curl of eq \abpi, provides the
actual link with string dynamics. Indeed, by projecting eq.\curlpi\ on the
world-sheet, one recovers the
equation of motion \lorentz\ for the ``~body~'' of the string. Instead,
eq.\loren\ describes the dynamics of the string end-points acted upon by the
``~internal force~'' $\p \dot X^\nu$, and by the ``~extended Lorentz force~''
$\left(f\kappa A_{\mu\nu}-eF_{\mu\nu}\right){\dot X}^\nu$. Thus,
eqs.(2.38)
and(2.40) are the equations that actually govern the dynamics of the string.

The other two field equations tell us something new: first
we note that eq.\dfj\ guarantees
that $H^{\mu\nu\rho}$ is a regular function. In fact, from equation (2.37)
$$\eqalign{
\partial_\nu\partial_\mu H^{\mu\nu\rho}&=
{f\kappa\over e}\partial_\nu\left(F^{\nu\rho}-{f\kappa\over
e}A^{\nu\rho}\right)
-f\kappa\partial_\nu J^{\nu\rho}\cr
&=f\kappa J^\rho-f\kappa J^\rho\equiv 0\ .\cr}
\eqn\consist
$$
Moreover, if one looks at
${f\kappa\over e}{\tilde A}^{\nu\rho}=
-F^{\nu\rho}+{f\kappa\over e}A^{\nu\rho}$ as a
``~gauge transformation~'' of $A^{\nu\rho}$, leaving the Kalb-Ramond field
strength $H^{\mu\nu\rho}$ unmodified, then eq.\dhj\ can be written as
a {\it London-type equation}
$$
\partial_\mu H^{\mu\nu\rho}-
m^2{\tilde A}^{\nu\rho}
=-\bar\kappa J^{\nu\rho}\ ,\qquad m^2\equiv {f\kappa\over e}\ ,
\qquad \bar\kappa\equiv f\kappa\ ,
\eqn\amass
$$
describing the propagation of a {\it massive, spin 1} field coupled to
its source $J^{\nu\rho}(x)$. So, the initially {\it massless, spinless} field
$A_{\mu\nu}$, because of its mixing with the vector gauge potential
$A_\mu$
acquires mass and spin: this is a peculiar mechanism through which
tensor gauge potentials become massive [\aashi]. Notice that the count of
degrees of freedom is the same as in the conventional Higgs mechanism:
a spin-1 gauge field with two degrees of freedom combines with a massless
spin-0 field thereby acquiring the
three degrees of freedom necessary to describe a massive spin-1 particle.
The difference here lies in the fact that the massless spin-0 field is itself
a gauge field, i.e., the rank-2 antisymmetric Kalb-Ramond
potential $A_{\mu\nu}$. On physical grounds,
equations (2.37-2.40) evoke the familiar picture of Abrikosov vortices
existing in type-2 superconductors. What is at work here is a
kind of relativistic Meissner effect: we have a massive
(superconducting) medium embedded into which are strings (vortices)
compressed by the pressure of the surrounding medium into lines of
electromagnetic flux. From here two scenarios come to mind.
The first is the familiar one of quark confinement (electric or magnetic)
originally advocated by Nambu: as flux tubes squeezed
by the pressure of a mass inducing medium,
such strings constitute very suitable traps for the hadron constituents
(string end points). Alternatively, one may be tempted to associate
such lines with ``~cosmic strings~'' as seeds of material structures in an
otherwise featureless sea of dark matter.\foot{A similar picture can be
constructed in terms of higher dimensional extended objects, i.e. membranes
or bags and higher rank antisymmetric tensor gauge fields [\five].}
In either scenario, eqs.(2.37) and (2.39) or eq. (2.42) apply to the
surrounding background medium, whereas equations (2.38) and (2.40) govern
the evolution of the string in such a background. Clearly, if there is any
element of truth in the above scenarios,
then two questions arise immediately: i) what is the effect of gravity on
the gauge formulation of string dynamics?, and ii) with an eye on the
Standard Model of particle physics, is it possible to attach internal
indices to the string geodesic field $\W$ ?

We will briefly examine both questions in the next section. However, for
completeness,
we shall close this section with the observation that the equations of
motion \dhj-\curlpi\ can be derived from the
effective action
$$\eqalign{
S^{\rm eff.}=&-\st\int_{\cal H}d^2\xi
\sqrt{-{1\over 2}{\dot X}_{\alpha\beta}{\dot X}^{\alpha\beta}}+\int
d^4x\left[
{f\kappa\over 2}\J(x)A_{\mu\nu}-{1\over 2\cdot
3!}H_{\mu\nu\rho}H^{\mu\nu\rho}
\right]\cr
&+\int d^4x\left[eA_\mu J^\mu-{1\over 4}
\left(F^{\nu\rho}+{f\kappa\over e}A^{\nu\rho}\right)
\left(F_{\nu\rho}+{f\kappa\over e}A_{\nu\rho}\right)\right]\cr
&-\mu_0\sum_{i=1,2}\int d\tau\sqrt{-{\dot X}_{i)}{}^\mu
{\dot X}_{i)\,\mu}}\cr}
\eqn\krop
$$
which is obtained by inserting the solution \solwj\ in the action \actg,
and represents an open string with massive charges at the end points.
``~Neutral~'' strings with massive end-points have been studied by several
authors [\chth],
[\bbhp], mainly in connection with hadron dynamics [\bh]. The novel feature
of $S^{\rm eff.}$ is the ``~ residual gauge symmetry~'' [\aashi]
$$\eqalign{
&\delta A_{\mu\nu}=\partial_{\,[\mu}\Lambda_{\nu]} \cr
&\delta A_\mu =\partial_\mu\phi
+{f\kappa\over e}\Lambda_\mu\ ,
\cr}
\eqn\extg
$$
surviving after the elimination of $B_{\mu\nu}$ in favor of the string
variables.

\chapter{Generally covariant and non-abelian formulation}

As we anticipated in the previous section, here we consider the possibility
of extending the formalism described so far in two directions: i) coupling
the system to gravity, and ii) non-abelian string geodesic fields.
\medskip
\underbar{\it Generally Covariant Formulation.}

For the sake of simplicity, we shall consider the interaction of closed
strings with the gravitational field. Accordingly, we substitute
in (2.1) the Minkowski metric with $g_{\mu\nu}(x)$, replace ordinary
derivatives with generally covariant ones $\nabla_\mu$, and add the
Einstein action. Thus, we are led to consider
$$\eqalign{
\sc=&-{\bar g}^2\int d^4x\sqrt{-g}\,\sqrt{-{1\over
2}g_{\mu\rho}g_{\nu\sigma}
W^{\mu\nu}W^{\rho\sigma}}+
{1\over 2}\int d^4x\sqrt{-g}\,
\W\nabla_{\,[\mu}B_{\nu]}\cr
&-{1\over 16\pi G}\int d^4x\sqrt{-g}\,R\cr
&\phantom{A}\cr
&\ff (x)\equiv \partial_{\,[\mu}B_{\nu]}(x)\equiv \nabla_{\,[\mu}B_{\nu]}\ .
\cr}
\eqn\covact
$$
Here, $g\equiv{\rm det}g_{\mu\nu}$.
The corresponding set of field equations represents the generally covariant
generalization of eqs.(2.2-2.3)
$$
\nabla_\mu\W=0\Rightarrow {1\over \sqrt{-g}}\partial_\mu\sqrt{-g}\W=0
\eqn\diwcov
$$
$$
-{\bar g}^2{\w\over\sqrt{-W^{\alpha\beta}W_{\alpha\beta}}}=
\bar F_{\mu\nu}\ ,\qquad
W^{\alpha\beta}W_{\alpha\beta}\equiv
{1\over 2}g^{\alpha\mu}g^{\beta\nu}W_{\alpha\beta}W_{\mu\nu}
\ ,\eqn\feqcov
$$
supplemented by the Einstein equations:
$$
R_{\mu\nu}-{1\over 2}g_{\mu\nu}R=8\pi G T_{\mu\nu}
\eqn\einst
$$
where the energy-momentum tensor in the r.h.s. is given by
$$
T_{\mu\nu}=-2\left[{\bar g^2\over 2}{W_{\mu\alpha}W_\nu{}^\alpha\over
\sqrt{-
W^{\alpha\beta}W_{\alpha\beta}}}+W_{\mu\alpha}F_\nu{}^\alpha\right]+
g_{\mu\nu}L.
\eqn\tmunu
$$
Then, a {\it closed} string appears as a special solution of \diwcov\ ,
namely
$$\eqalign{
&\hat\W(x)={c\over \sqrt{-g}}\int_{\cal H}d^2\xi\,\delta^{4)}\left(x-
X(\xi)\right){\dot X}^{\mu\nu}=
c\,J^{\mu\nu}(x)\ ,\cr
&c= \hbox{dimensionless const.}\ ,\cr}\eqn\covsol
$$
which is nothing but the general covariant form of the string-current.
To demonstrate the equivalence with gravity coupled to a closed string we
still
have to show that \einst\ is the Einstein field equation with a string source
in the r.h.s. To do that, we notice that the Lagrangian in \tmunu\ vanishes
on-shell, i.e. $\displaystyle{L(\bar F;W)=0}$. Then
$$\eqalign{
T_{\mu\nu}&=
\bar g^2 c^2{J_{\mu\alpha}J_\nu^\alpha\over
\sqrt{-{1\over 2}J^{\alpha\beta}J_{\alpha\beta}}}\cr
&={\bar g^2 c^2\over\sqrt{-g}}
\int d^3\xi{{\dot X}_{\mu\alpha}{\dot X}_\nu{}^\alpha\over
\sqrt{-{1\over 2}{\dot X}^{\beta\gamma}{\dot X}_{\beta\gamma}}}
\delta^{4)}\left(x-X(\xi)\right)\ ,\cr}
\eqn\tstring
$$

and the equivalence with General Relativity becomes manifest once we show
that the energy-momentum tensor has vanishing divergence This is indeed
the case if $X^\mu(\xi)$ represents a classical solution of the
string equation of motion. In fact

$$\eqalign{
\nabla_\mu T^{\mu\nu}&=
{\bar g^2 c^2\over\sqrt{-g}}
\int d^2\xi{{\dot X}^\mu{}_\alpha{\dot X}^{\nu\alpha}\over
\sqrt{-{1\over 2}{\dot X}^{\beta\gamma}{\dot X}_{\beta\gamma}}}
\delta^{4)}\nabla_\mu\left(x-X(\xi)\right)\cr
&=-{\bar g^2 c^2\over\sqrt{-g}}
\int d^2\xi\delta^{[ma]}{\partial_a X_\alpha {\dot X}^{\nu\alpha}\over
\sqrt{-{1\over 2}{\dot X}^{\beta\gamma}{\dot X}_{\beta\gamma}}}
\nabla_m\delta^{4)}\left(x-X(\xi)\right)\cr
&=-{\bar g^2 c^2\over\sqrt{-g}}\int d^2\xi\delta^{[ma]}
\Biggl[\nabla_m\left({\partial_a X_\alpha {\dot X}_{\nu\alpha}\over
\sqrt{-{1\over 2}{\dot X}^{\beta\gamma}{\dot X}_{\beta\gamma}}}
\delta^{4)}\left(x-X(\xi)\right)\right)\cr
&-\nabla_m\left({\partial_a X_\alpha {\dot X}^{\nu\alpha}\over
\sqrt{-{1\over 2}{\dot X}^{\beta\gamma}{\dot X}_{\beta\gamma}}}\right)
\delta^{4)}\left(x-X(\xi)\right)\Biggr]\ .\cr}
\eqn\divt
$$
The first term in the last line of equation\divt\ is a pure surface integral,
which is zero for a
closed string. Furthermore, anti-symmetrized
covariant derivatives can be replaced with ordinary
partial derivatives, and this yields the desired result,
$$\eqalign{
\nabla_\mu T^{\mu\nu}&={\bar g^2 c^2\over\sqrt{-g}}\int
d^3\xi\delta^{[ma]}
\partial_m\left({\partial_a X_\alpha {\dot X}^{\nu\alpha}\over
\sqrt{-{1\over 2}{\dot X}^{\beta\gamma}{\dot X}_{\beta\gamma}}}\right)
\delta^{4)}\left(x-X(\xi)\right)\cr
&={1\over\sqrt{-g}}\int d^3\xi\left[\delta^{[ma]}
\partial_m\Pi^{\nu\alpha}\partial_a X_\alpha\right]
\delta^{4)}\left(x-X(\xi)\right)=0\ .\cr}
\eqn\divtwo
$$
{\it Notice how the consistency condition (3.9) immediately gives the
equation of motion of the string as an alternative to projecting the Bianchi
identity for $\bar F_{\mu\nu}$ on the string}
world-sheet. This property represents a distinct advantage of the generally
covariant formulation of our model. Ultimately, of course, the consistency
condition for the
Einstein equations coupled to matter, i.e., that both the Ricci tensor and the
energy
momentum tensor must have vanishing covariant divergence, can be traced
back to the Bianchi Identity for the Riemann tensor.

At this point one might elect to investigate the nature of the solutions of
the classical system (3.1). However, in view of the equivalence that we have
just established, a discussion of some such solutions already exists in the
literature, especially
in connection with cosmic strings [\sixten]. Presently, what interests us is
a deeper
conceptual question: with an eye on quantum gravity and on its attending
difficulties, is it really necessary to include from the very beginning the
Einstein term in the action (3.1)?

We are partly led to this question by the result communicated in a
previous article [\new] where we have shown that General Relativity may
arise
as the low energy limit of a quantum theory of relativistic {\it membranes}.
Indeed, the current attitude
towards ultra short distance physics is to replace local fields with
extended
objects, mainly strings, as fundamental constituents of matter and to treat
particle physics below some (string) energy scale as a local limit of the
fundamental theory. In the
following we shall argue briefly that our gauge formulation of string
dynamics is perfectly consistent with this point of view.

Suppose we start from the action
$$\eqalign{
\sc=&-{\bar g}^2\int d^4x\sqrt{-g}\,\sqrt{-{1\over
2}g_{\mu\rho}g_{\nu\sigma}
W^{\mu\nu}W^{\rho\sigma}}+
{1\over 2}\int d^4x\Bigl[\sqrt{-g}\,
\W\nabla_{\,[\mu}B_{\nu]}\cr
&+\W K_{\mu\nu}\Bigr]\ ,\cr}
\eqn\newact
$$
where $g_{\mu\nu}(x)$ and  $K_{\mu\nu}(x)$ represent respectively a
symmetric
and an anti-symmetric
arbitrary external source, i.e., they are background fields
implementing invariance under general coordinate transformations and
extended
gauge invariance,
$$\eqalign{
&g'_{\mu\nu}(x')={\partial x^\rho\over\partial x^{'\,\mu}}
{\partial x^\sigma\over\partial x^{'\,\nu}}
g_{\rho\sigma}(x)\ ,\cr
&W^{'\,\mu\nu}(x')={\partial x^{'\,\rho}\over\partial x^\mu}
{\partial x^{'\,\sigma}\over\partial x^\nu}
W^{\rho\sigma}(x)\ ,\cr
&B'_\mu(x')={\partial x^\rho\over\partial x^{'\,\mu}}B_\rho(x)\ ,\cr
&\delta B_\mu(x)=\Lambda_\mu(x)+\partial_\mu\phi(x)\ ,\cr
&\delta K_{\mu\nu}(x)=\partial_{\,[\mu}
\Lambda_{\nu]}\ .\cr}
\eqn\newinv
$$
At this stage  there is no relationship between $g_{\mu\nu}(x)$ and the
physical spacetime metric, nor between
$K_{\mu\nu}(x)$ and the Kalb-Ramond tensor potential. However, once
$\W(x)$
is eliminated
from \newact\ by means of the formal solution \sol, we obtain the action
for a string non-linear $\sigma$-model [\call]:
$$
\sc=-{\bar g}^2c\int d^2\xi\,\sqrt{-{1\over
2}g_{\mu\rho}(X)g_{\nu\sigma}(X)
{\dot X}^{\mu\nu}{\dot X}^{\rho\sigma}}+
{c\over 2}\int d^2\xi\, K_{\mu\nu}(X){\dot X}^{\mu\nu}\ .
\eqn\sigmod
$$
Then, an effective action for the background fields is induced at
the two-loop quantum level [\kik], and can be computed in a perturbative
expansion in powers of the inverse effective string tension:
$$
\Gamma=-{\bar g}^2c\int d^4x\sqrt{-g}\left[a g^{\mu\nu}R_{\mu\nu}-d
H^{\mu\nu\rho}H_{\mu\nu\rho}+O\left[\left(1/{\bar
g}^2c\right)^{3/2}\right]
\right]\ .
\eqn\effact
$$
Thus, in such an approach, the Einstein and Kalb-Ramond terms are
recovered as
induced quantum terms, describing the low-energy behavior of the
underlying quantum string theory.
\medskip

\underbar{\it Non-Abelian geodesic field}

Spatially extended objects were introduced into hadronic physics after the
recognition that the energy spectrum of the dual resonance model could be
interpreted in terms of the vibrating modes of a relativistic string [\dual].
Further elaboration of this idea led to models of the meson as a pair of
colored quarks, or monopoles,
joined by a thin flux-tube [\namb,\eguchi].

Therefore, if we believe that the geodesic field approach discussed above is
general
enough to provide a consistent description of elementary strings as well as
gauge strings, then it should be possible to embody
non-abelian symmetries into the proposed approach. It turns out, however,
that there are severe restrictions on the feasibility of this program.
The rest of this section is devoted to discuss this problem.

Again, we shall follow a line of reasoning similar to the one proposed in the
previous section, i.e., we start from the non-abelian current associated with
a pair
of ``~colored~'' point-like objects\foot{What we have in mind is the $QCD$
string, so we
consider $SU(3)_{\rm c}$ as the underlying symmetry group of strong
interactions, and let the color index  $a=1,\dots\ 8$.

In order to distinguish internal indices we shall write them always as
upper indices. Repeated indices are traced over with an euclidean metric.}
$$
J^{a\,\mu}=\int d\tau \sum_{i=1,2}(-1)^i\delta^{4)}
\left(x-X_i(\tau)\right)\rho^a\left[X_{i)}(\tau)\right]
\left({dX_i^\nu\over d\tau}\right)\ ,
\eqn\jq
$$
which acts as the source of an $SU(3)$ gauge field $A^a{}_\mu$ governed by
the Yang-Mills action.  $\rho^a[X]$ is the a-th component of the Yang-Mills
charge carried by the ``~quark~'' located at $x^\mu=X_i^\mu(\tau)$.

By implementing again the ``~Dirac trick~'',
we write $J^{a\,\mu}$ as the gauge covariant divergence
of a singular Yang-Mills field
$$
G^{a\,\mu\nu}(x)=\int d\tau d\sigma\rho^a[X]\delta^{4)}\left(x-X(\xi)\right)
{\dot X}^{\mu\nu}
\ ,
\eqn\js
$$
where $x^\mu=X^\mu(\tau,\sigma)$ describe a world-sheet having the two
world-lines $x^\mu=X_{i)}{}^\mu(\tau)$ as its only boundary, i.e.
$$
x^\mu=X^\mu\left(\tau,\sigma=\sigma_i(\tau)\right)=X^\mu_i(\tau)\ ,\quad
i=1,2\ .
\eqn\bordi
$$

In order to implement the
Dirac relationship between $G^{a\,\mu\nu}$ and $J^{a\,\mu}$ we
require that $\rho^a[X]$ be covariantly constant [\rysh], i.e.
$$
D^{ab}{}_\mu\rho^b[X]=0\Rightarrow D^{ab}{}_\mu G^{b\,\mu\nu}=J^{a\,\nu}\ .
\eqn\current
$$
This constraint is reasonable since $\rho^b[X]$ is not a dynamical variable,
but rather an external source which can be suitably chosen. It also follows
from \current\ that $\displaystyle{(\rho^a)^2\equiv \rho^a\rho^a }$
is independent of the world-sheet coordinates
$$
\partial_\tau (\rho^a)^2=\partial_\sigma (\rho^a)^2=0\ .
\eqn\indep
$$
It is also worthwhile to remark that the regularized form of
$G^{a\,\mu\nu}$
evaluated on the string world-sheet factorizes in the product of the
color distribution times the (~{\it abelian}~) volume momentum
$$
\eqalign{
G^{a\,\mu\nu}(x=X)&=
\int d^2\xi'\rho^a[Y]\delta^{4)}\left(X(\xi)-Y(\xi')\right)\dot Y^{\mu\nu}\cr
&={\rm const.}{1\over a^2}\int d^2\xi'\rho^a[Y]\delta^{2)}\left(\xi-
\xi'\right)
{\dot X^{\mu\nu}\over\sqrt{-{1\over2}{\dot X}_{\rho\sigma}
{\dot X}^{\rho\sigma}}}\cr
&={\rm const.}'\rho^a[X]\Pi^{\mu\nu}(\xi)\ .\cr}
\eqn\sing
$$
Accordingly, we establish the following relation between the string field
$G^{a\,\mu\nu}(x) $ and the slope field:
$$
G^{a\,\mu\nu}(x)=
\SLF(x)\int d^2\xi\rho^a[X]\,\delta^{4)}\left(x-X(\xi)\right)\ .
\eqn\cslope
$$
{}From the above equation one derives  a formal expression for
$G^{a\,\mu\nu}$
squared:
$$\eqalign{
G^{a\,\mu\nu}(x)G^a{}_{\mu\nu}(x)&=
\SLF(x)\slf(x)\int d^2\,\xi d^2\xi'\rho^a[X]\rho^a[X]\,
\delta^{4)}\left(x-X(\xi)\right)\delta^{4)}\left(x-X(\xi')\right)\cr
&=\SLF(x)\slf(x)(\rho^a)^2\left(\int d^2\,\delta^{4)}\left(x-X(\xi)\right)
\right)^2\ ,\cr}
\eqn\jsquare
$$
which we shall use later on.

The above remarks suggest we can write an $SU(3)$ invariant action for the
string geodesic field as follows
$$
\eqalign{
{\cal S}^{\rm c.}=&\int d^4x\Biggl[-{\bar g}^2\sqrt{-{1\over 3!}
W^a{}_{\mu\nu}W^{a\,\mu\nu}}+
{1\over 2}W^{a\,\mu\nu}(x)D^{ab}{}_{\,[\mu}B^b{}_{\nu]}(x)
+fB^a{}_\mu(x)J^{a\,\mu(x)}\cr
&+eA^a{}_\mu J^{a\,\mu}-{1\over 4}F^a{}_{\mu\nu}F^{a\,\mu\nu}\Biggr]
-\mu_0\int d\tau\sum_{i=1,2}\sqrt{-{\dot X}_{i)}{}^\mu{\dot X}_{i)\,\mu}}
\ ,\cr}
\eqn\gact
$$
where $J^{a\,\mu}(x)$ is given by \jq\ and $D^{ab}{}_{\,[\mu}B^b{}_{\nu]}(x)$
is the gauge covariant derivative of the $B$-field.
The colored objects in \gact\ transform under color internal rotations as
$$\eqalign{
&\delta W^{a\,\mu\nu}(x)=f^a{}_{bc}\Lambda^b W^{c\,\mu\nu}(x)\ ,\cr
&\delta B^a{}_\mu(x)=D^{ab}{}_\mu\Lambda^b\ ,\cr
&\delta\rho^a[X]=f^a{}_{bc}\Lambda^b\rho^c[X]\ ,\cr}
\eqn\ymtransf
$$
where $f^a{}_{bc}$ are the $SU(3)$ structure constants.
The corresponding field equations are
$$
\delta_{B_{\mu}}{\cal S}^{\rm c.}=0:
\qquad D^{ab}{}_\mu G^{b\,\mu\nu}(x)=fJ^{a\,\nu}(x)\ ,
\eqn\gdwj
$$
$$
\delta_{W_{\mu\nu}}{\cal S}^{\rm c.}=0:\qquad
{\bar g}^2{G^a{}_{\mu\nu}(x)\over\sqrt{-{1\over 2}G^b{}_{\alpha\beta}
G^{b\,\alpha\beta}}}
+D^{ac}{}_{\,[\mu}B^c{}_{\nu]}=0\ ,\eqn\gabw
$$
$$
\delta_{X_{i)}{}^{\mu}(\tau)}{\cal S}^{\rm c.}=0:\qquad
{dP_{i)\,\mu}\over d\tau}=
(-1)^i\rho^a[X_{i)}]\left(f D^{ab}_{\,[\mu}B^b{}_{\nu]}+
eF^a{}_{\mu\nu}\right)
{\dot X}_{i)}{}^\nu\ , \eqn\gpdot
$$
$$
\delta_{A_{\mu}}{\cal S}^{\rm c.}=0:\qquad D^{ab}_\mu
F^{b\,\mu\nu}=eJ^{a\,\nu}
\ . \eqn\gmax
$$
{}From the first two equations we derive the following formal solutions
in terms of string variables
$$
W^{a\,\mu\nu}(x)=fG^{a\,\mu\nu}(x)\quad\Rightarrow
W^{a\,\mu\nu}(x=X)={\rm const.}\rho^a[X]\Pi^{\mu\nu}(\xi)\ ,
\eqn\solone
$$
and
$$
B^a{}_{\mu\nu}(x)\equiv D^{ac}{}_{\,[\mu}B^c{}_{\nu]}=-
{\bar g}^2{G^a{}_{\mu\nu}(x)\over\sqrt{-{1\over 2}G^b{}_{\alpha\beta}
G^{b\,\alpha\beta}}}\quad\Rightarrow-{1\over 2}B^a{}_{\mu\nu}(x)
B^{a\,\mu\nu}(x)={\bar g}^2\ .
\eqn\soltwo
$$
Equation \soltwo\ allows us to introduce a colored geodesic field
$$
\Pi^a{}_{\mu\nu}(x)\equiv -fB^{a\,\mu\nu}(x)
\eqn\cpi
$$
which on the string world-sheet reads
$$
\Pi^a{}_{\mu\nu}(x=X)={{\bar g}^2 f\over\sqrt{(\rho^a)^2}}
{\rho^a[X]{\dot X}_{\mu\nu}\over
\sqrt{-{1\over 2}{\dot X}_{\mu\nu}{\dot X}^{\mu\nu}}}
\equiv\st
{\rho^a[X]{\dot X}_{\mu\nu}\over
\sqrt{-{1\over 2}{\dot X}_{\mu\nu}{\dot X}^{\mu\nu}}}\ ,
\eqn\projpi
$$
and represents the area-momentum of a colored string with an effective
tension
$$
\st\equiv {{\bar g}^2 f\over\sqrt{(\rho^a)^2}}\ .
\eqn\efft
$$
{}From \soltwo, \cpi\ and the constraint \current, we deduce that the
color-singlet
$\displaystyle{\rho^a[X]B^{a\,\mu\nu}(x)}$ satisfies the Bianchi Identity
$$
\partial_{[\,\lambda}\rho^a B^a{}_{\mu\nu]}=0\ .
\eqn\cbianchi
$$
Once evaluated along the world-sheet, eq.\cbianchi\ yields the
equations of motion for $X(\tau,\sigma)$:
$$
{\dot X}^{\lambda\mu}\partial_{[\,\lambda}\rho^a[X]B^a{}_{\mu\nu]}(\xi)=0\
,
\Rightarrow
{\bar g}^2\sqrt{(\rho^a)^2}\delta^{[jk]}\partial_j X^\mu\partial_k
{{\dot X}_{\mu\nu}\over\sqrt{-{1\over 2}{\dot X}_{\mu\nu}{\dot
X}^{\mu\nu}}}
=0\ .
\eqn\cstring
$$
Therefore, the system of eqs. \gdwj-\gmax can now be written in the
familiar ``~string~'' form:
$$
\delta^{[jk]}\partial_j X^\mu
\partial_k\Pi_{\mu\nu}(\xi)=0\ ,
\eqn\ops
$$
$$
{dP_{i)\,\mu}\over d\tau}=
(-1)^i\left(\Pi_{\mu\nu} +
e\rho^a[X_{i)}]F^a{}_{\mu\nu}\right)
{\dot X}_{i)}{}^\nu\ , i=1,2 \eqn\clorentz
$$
$$
D^{ab}_\mu F^{b\,\mu\nu}=eJ^{a\,\nu}
\ . \eqn\ym
$$
The first equation of this system is nothing but \cstring\ written by
taking into
account eq.\efft, and describes a minimal surface
traced in spacetime by the evolution of the body of a string.
Equations\clorentz\ represent a generalization of the Wong equation
[\wong],
and describe
the motion of the string end-points under the influence of both the
Yang-Mills field and the ``~internal Lorentz force~'' represented by $\p$.
Finally, the colored string boundary enters eq.\ym\ as the source of the
Yang-Mills field itself. The whole system of eqs.\ops-\ym\ can be derived
through variation of the effective action functional obtained by inserting
\solone,\soltwo\ into \gact :
$$\eqalign{
{\cal S}^{\rm eff.}&=\int d^4x\Biggl[-{\bar g^2 f}\sqrt{(\rho^a)^2}
\sqrt{-{1\over2}\slf(x)\SLF(x)}\int d^2\xi\,\delta^{4)}\left(x-
X(\xi)\right)\cr
&+eA^a{}_\mu J^{a\,\mu}-{1\over 4}F^a{}_{\mu\nu}F^{a\,\mu\nu}\Biggr]
-\mu_0\int d\tau\sum_{i=1,2}\sqrt{-{\dot X}_{i)}{}^\mu{\dot
X}_{i)\,\mu}}\cr
&=-\int d\tau\left[\st\int d\sigma
\sqrt{-{1\over 2}{\dot X}_{\mu\nu}{\dot X}^{\mu\nu}}+
\mu_0\sum_{i=1,2}\sqrt{-{\dot X}_{i)}{}^\mu{\dot X}_{i)\,\mu}}\right]\cr
&\>\>-{1\over 4}\int d^4x\, F^a{}_{\mu\nu}F^{a\,\mu\nu}
+e\int d\tau \sum_{i=1,2}(-1)^i\rho^a\left[X_{i)}\right]
{\dot X}^\mu{}_{i)}A^a{}_\mu(X_{i)})\ .
\cr}
\eqn\sceff
$$
The functional \sceff\ is a generalization of the ``~massive ends~''
string action suggested by Chodos and Thorne [\chth], including color and
interacting with a Yang-Mills field.

{\it Notice how the color degree of freedom disappear from the Nambu-Goto
action,
having been completely re-absorbed into the definition of the string
tension.}
In fact, non-abelian gauge symmetry is incompatible with
reparametrization invariance for any kind of spatially extended
object [\teitel], so that a true realization of a colored string seems to
be impossible. In this connection, observe that in eqs.(3.17) and (3.29),
the color charge $\rho^a[X]$ simply
multiplies the ordinary(=abelian) string variables, and thus
 remains localized at the pointlike string boundaries.
Accordingly, the coupling of the string with the Yang-Mills field occurs
{\it only} along the world-lines of the pointlike boundaries.

Now, it would be tempting to inquire if a \KR\ interaction
can be added in a sensible way, or in other
words, if the gauge transformations (2.27) admit a  non-abelian
generalization.
The main problem is to introduce a suitable kinetic term for the internal
algebra valued Kalb-Ramond tensor potential $A^a{}_{\mu\nu}(x)$. In fact,
the naive extension $H^{a\,\mu\nu\rho}H^a{}_{\mu\nu\rho}$ breaks the
vector
gauge invariance. Possible solutions to this problem are
currently under investigation along two main lines of thought: one is to
implement  the equivalence of a certain class of non-abelian
Kalb-Ramond theories and chiral, non-linear $\sigma$-models [\freetow];
the
second consists in compensating the lack of vector gauge symmetry by
means
of suitable matter fields [\smol].   The final goal is to introduce a new mass
generating mechanism  arising from the mixing between
tensor and vector gauge bosons.

\refout
\bye